\documentclass[aps,pra,twocolumn,a4paper,groupedaddress,floatfix,10pt,longbibliography]{revtex4-1}
\usepackage{graphicx,amsmath,amssymb,amsfonts,dsfont,enumitem,MnSymbol,comment,array,tabularx}
\newcolumntype{P}[1]{>{\centering\arraybackslash}p{#1}}
\newcommand{\mf}[1]{\boldsymbol{#1}}

\newcommand{\mc}[1]{\ensuremath{\mathcal{#1}}}

\usepackage{color,ulem}

\newcommand{\Ns}{N}

\newcommand{\sm}{d}

\newcommand{\lb}{A}
\newcommand{\md}{B}
\newcommand{\rb}{C}
\newcommand{\mps}{M} 

\newcommand*{\vcenteredbox}[1]{\begin{tabular}{@{}l@{}}#1\end{tabular}}
\begin{document}
   
\title{
Tensor network reduced order models for wall-bounded flows 
}
\author{Martin Kiffner${}^{1}$}
\author{Dieter Jaksch${}^{2,1}$} 
\affiliation{Clarendon Laboratory, University of Oxford, Parks Road, Oxford OX1
3PU, United Kingdom${}^1$} 
\affiliation{Institut für Laserphysik, Universit\"{a}t Hamburg, Hamburg, Germany${}^2$}

\begin{abstract}
We introduce a widely applicable tensor 
network-based framework for developing 
reduced order models describing wall-bounded 
fluid flows. 
As a 
paradigmatic example, we consider 
the  incompressible Navier-Stokes equations 
and the lid-driven cavity in two spatial dimensions. 
We benchmark our solution against published reference data for low Reynolds numbers and find excellent 
agreement. In addition, we investigate  the  
short-time dynamics of the flow 
at high Reynolds numbers for the lid-driven 
and doubly-driven cavities. 
We represent the  velocity  
components by matrix product states 
and find that the bond dimension grows logarithmically with 
simulation time. 
The tensor network algorithm requires at most 
a few percent  of 
the number of variables parameterizing the 
solution obtained by direct numerical simulation, 
and approximately improves the runtime by 
an order of magnitude compared to direct numerical simulation on similar hardware. 
Our approach is readily 
transferable to other flows, 
and paves the 
way towards quantum computational fluid 
dynamics in complex geometries.
\end{abstract}

\maketitle

\section{Introduction \label{intro}}
Direct numerical simulation (DNS) of the 
Navier-Stokes equations at large Reynolds 
numbers would be a highly desirable capability 
for science and engineering applications. 
However, it remains 
an elusive goal  due to the extremely large 
numerical  complexity associated with the 
multiscale nature 
of turbulence~\cite{monin:71,monin:75}. 
The state-of-the-art method for  mitigating 
this issue in computational fluid 
dynamics (CFD) is turbulence 
modelling~\cite{hanjali:mt}, which continues to be under 
constant development for improving its accuracy.

A conceptionally different approach  
to reducing the numerical complexity of DNS 
is through structure-resolving methodologies~\cite{HLB96,AIAA17}. 
These methods aim to establish a reduced order 
model (ROM) of the full system  by exploiting correlated structures in the solution.
However, identifying suitable modes for 
building ROMs is difficult and therefore  
under active investigation~\cite{AIAA17,ddse:19,kutz:dmd,ramezanian:21}.

Recently,  quantum-inspired tensor network methods have been introduced as  a novel paradigm 
for modelling turbulent flows for diagnostic and predictive purposes~\cite{gourianov:22}. 
The tensor network  algorithm in~\cite{gourianov:22} 
for solving the incompressible Navier-Stokes equation (INSE)
approximates the velocity components in
matrix product state (MPS) format~\cite{schollwoeck:11}. In 
the examples studied in~\cite{gourianov:22}, 
the number of variables parameterizing the 
solution (NVPS) in MPS representation
is reduced by over an order of 
magnitude  
compared to DNS. The MPS algorithm 
thus realizes a 
ROM for the  investigated flows. 
However, the efficient compression reported 
in~\cite{gourianov:22} only resulted 
in a computational speedup in a one-dimensional 
system, but not in  two or 
three spatial dimensions. Furthermore, the examples in~\cite{gourianov:22} are restricted to homogeneous flows with periodic boundary conditions. 
Here we show that ROMs  based on 
tensor networks can be extended to 
wall-bounded flows. We illustrate our approach 
using the lid-driven cavity in two 
spatial dimensions, which is a very well-studied 
problem~\cite{shankar:00} with tabulated reference 
solutions~\cite{ghia:82}. 
We solve the INSE in the streamfunction-vorticity formulation and find that our MPS algorithm reproduces the data in~\cite{ghia:82} for stationary states at low Reynolds numbers. 

As an application of our approach, we 
assume that the  fluid 
is initially at rest and  
investigate the short-time dynamics 
at high Reynolds numbers. We represent 
the  velocity 
components by MPSs with bond dimension $\chi$ 
and investigate how $\chi$ depends on time and
grid size. We find that 
$\chi$ grows logarithmically in time 
and reduces the NVPS
compared to direct numerical simulation by about 97\%.

As an extension towards more complex flows, 
we also investigate the doubly-driven cavity 
where both the top and bottom lids move and 
find the same qualitative behavior. We 
compare the runtimes of the MPS and DNS algorithms
on similar hardware and at different Reynolds numbers. We find that the MPS algorithm 
can give rise to significant runtime improvements 
compared to DNS, peaking at a seventeen-fold 
speedup in case of the lid-driven cavity.

The MPS algorithm in~\cite{gourianov:22} 
advances the solution to the INSE by solving an 
optimization problem. More specifically, 
the continuity equation is combined 
with the momentum equations via the 
penalty method~\cite{rider:05}, and 
the updated velocity components are 
obtained by minimizing 
a single cost function. 
On the contrary,  the MPS algorithm in 
this work is constructed by emulating 
the DNS algorithm step-by-step. 
We achieve this by decomposing 
the DNS algorithm into four 
elementary operations (multiplication, addition, 
matrix-vector operations and solving 
linear systems of equations) that can be realized in MPS format. It follows 
that our approach is directly transferable 
to a broad class of other CFD  methodologies and flow geometries. 

An important feature of quantum-inspired tensor network algorithms 
is that they can be ported to a quantum 
computer~\cite{gourianov:22,fukagata:22}.  
This transfer can be achieved 
with quantum circuits of known  depth~\cite{lubasch:20} and 
will provide at least a  quadratic speedup over 
the scaling of the classical tensor network algorithm 
with the bond dimension~\cite{gourianov:22}. Improved speedups may be achieved by 
problem-specific quantum circuits~\cite{lubasch:20,Pollman2020,Pollman2021} that perform exponentially better than 
the MPS  encoding of flow fields. 
Our work thus represents a first step towards efficient quantum 
algorithms for solving CFD 
problems with boundary conditions.

This paper is organized as follows. 
The model for the lid-driven cavity in the 
streamfunction-vorticity formulation is presented in Sec.~\ref{model}. We give a 
detailed description of the model and the 
spatial discretization because this forms 
the foundation for constructing the MPS 
algorithm. We outline the encoding of 
flow fields in MPS format and describe 
how the DNS algorithm can be transformed 
into MPS format. All technical details 
are summarized in Appendices. 
The  results are shown in Sec.~\ref{results} 
and begin with a validation of our tensor network 
algorithm against previous work. We then consider 
the short-time dynamics following the quench by the 
moving lid and analyze the bond dimension as a function 
of time and grid size. 
A summary and discussion of our results is provided in Sec.~\ref{summary}. 

\section{Model \label{model}} 
\begin{figure}[t!]
\begin{center} 
\includegraphics[width=\columnwidth]{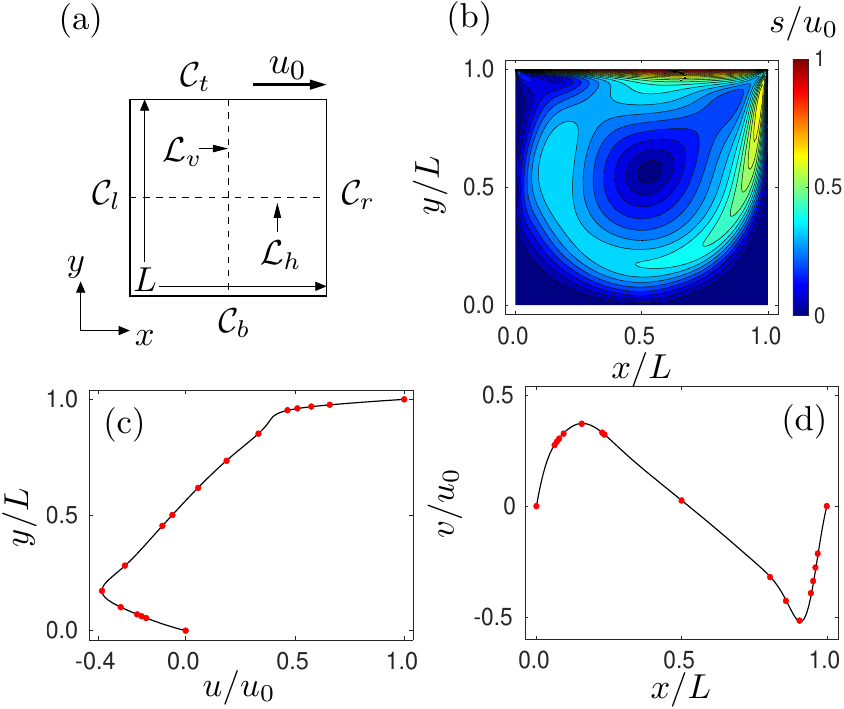}
\end{center}  
\caption{\label{fig1} 
(Color online) (a) Setup of the square lid-driven cavity with edge length $L$. 
The upper lid moves at constant velocity $u_0$ 
in $x$-direction. $\mc{C}_t$, 
$\mc{C}_r$, $\mc{C}_b$, $\mc{C}_l$ are the  top, right, bottom and left boundaries, respectively. $\mc{L}_v$ 
($\mc{L}_h$) denotes a vertical (horizontal) line 
through the center of the cavity.
(b) Contour plot of the velocity magnitude  $s=\sqrt{u^2+v^2}$ at $t=50$ for  Re=1000 and evaluated with the tensor network algorithm.
(c) Comparison of the tensor network solution 
for the $x$-component $u$ of the velocity 
along $\mc{L}_v$ (black solid line) 
with the reference values in~\cite{ghia:82} 
(red dots). 
(d) Comparison of the tensor network solution 
for the $y$-component $v$ of the velocity 
along $\mc{L}_h$ (black solid line) 
with the reference values in~\cite{ghia:82} 
(red dots). 
}
\end{figure}

The setup for the lid-driven cavity in 
two spatial dimensions is shown in 
Fig.~\ref{fig1}(a). 
We consider a square box with edge length $L$, 
and the upper lid moves 
with velocity $u_0$ in $x$-direction. 
The $x$ component ($y$ component) of the 
fluid is denoted by $u$ ($v$). At $t=0$, 
the fluid is at rest, $u=v=0$. We consider 
a viscous fluid with kinematic 
viscosity $\nu$ and seek solutions to the 
incompressible Navier-Stokes equations 
in the streamfunction-vorticity 
approach~\cite{rider:05}, 
\begin{subequations}
 \begin{align}
  \partial_t w & = -\left[\partial_x(u w) + \partial_y(v w)\right] + \nu \Delta w\,,
  \label{momentum} \\
    \Delta \psi & = -w \, .\label{poisson}
 \end{align}
 \label{eqstosolve}
\end{subequations}
The  streamfunction $\psi$ and the velocity 
components $u$ and $v$ are connected via
\begin{subequations}
 \begin{align}
  u & = \partial_y\psi\,,\\
  v & = -\partial_x \psi\,,
 \end{align}
\label{velocities}
\end{subequations}
where
\begin{equation}
 w=\partial_x v - \partial_y u
\end{equation}
is the vorticity. Throughout this work 
we scale time in units of $t_0=L/u_0$, length 
in terms of $L$ and velocities by $u_0$. 
Solutions to Eq.~(\ref{eqstosolve}) are 
then characterized by the Reynolds number
\begin{align}
 \text{Re} = \frac{u_0 L}{\nu}\,.
\end{align}
\begin{table}[b!]
 
\centering

\begin{tabular}{c||c|c|c|c|}
 & $\mc{C}_t$ & $\mc{C}_r$ & $\mc{C}_b$ & $\mc{C}_l$\\
 \hline
 \hline
 $u$ & $u_0$ & 0 & 0 & 0 \\
 \hline
 $v$ & 0 & 0 & 0 & 0 \\
 \hline
 $\psi$ & 0 & 0 & 0 & 0 \\
 \hline 
\end{tabular}
\caption{\label{tabB} 
Dirichlet boundary conditions for 
velocity fields $u$, $v$ and the 
streamfunction $\psi$ on 
boundaries $\mc{C}_{\alpha}$ as indicated in 
Fig.~\ref{fig1}(a).}
\end{table}
%
%
%
We discretize the interior of the cavity (excluding 
boundaries) by a uniform grid with 
$K$ grid points in each spatial dimension. 
The computational domain thus comprises $K^2$ equally spaced 
points $\mf{r}_k$ with grid spacing 
\begin{align}
h=L/(K+1)\,.
\end{align}
Each grid point vector 
$\mf{r}_k$ is uniquely described by 
a tuple of integers, 
\begin{equation}
 \mf{r}_k\leftrightarrow(k^x,k^y)\,,
 \label{onemap}
\end{equation}
where $k^{\alpha} \in\{0,\ldots,K-1\}$ is the index of the grid point in the direction $\hat{\mf{e}}_{\alpha}$ with $\alpha\in\{x,y\}$.
The one-to-one correspondence in Eq.~(\ref{onemap}) allows us 
to label discrete 
function values on the grid by
$F(\mf{r}_k) \equiv F_{k^x,k^y}$. We denote 
ghost points on the left (bottom) boundary 
by $k^x=-1$ ($k^y=-1$), and those on the right 
(top) boundary by $k^x=K$ ($k^y=K$). 
The streamfunction $\psi$ must vanish everywhere on the boundary, and all velocity 
components are zero except for $u=u_0$ 
on boundary $\mc{C}_t$ [see Fig.~\ref{fig1}(a)]. 
The boundary conditions for $\psi$, $u$ and 
$v$ are summarized in Tab.~\ref{tabB}. 
We obtain the boundary values for 
the vorticity $w$ in the standard 
approach~\cite{ghia:82} and find 
($p,q\in\{0,\ldots,K-1\}$) 
\begin{subequations}
 \begin{align}
 w_{p,K} & = -\frac{3}{h}u_0 
 +\frac{1}{h^2}\left(- 4\psi_{p,K-1} + \frac{1}{2}\psi_{p,K-2} \right)\,,\label{uterm}\\
 w_{p,-1} & = \frac{1}{h^2}\left(- 4\psi_{p,0} + \frac{1}{2}\psi_{p,1} \right) \,,\\
 w_{-1,q} & = \frac{1}{h^2}\left(- 4\psi_{0,q} + \frac{1}{2}\psi_{1,q} \right) \,,\\
w_{K,q} & = \frac{1}{h^2}\left(- 4\psi_{K-1,q} + \frac{1}{2}\psi_{K-2,q} \right) \,.
 \end{align}
\label{boundaryw}
\end{subequations}
%
%
The DNS algorithm for solving Eq.~(\ref{eqstosolve}) 
with the boundary conditions in Tab.~\ref{tabB} 
is outlined in Appendix~\ref{dns}. 
For the time integration of Eq.~(\ref{momentum}), 
we use a second-order 
MacCormack algorithm~\cite{MacCormack:69,rider:05,anderson:cfd}. 
Finite-difference operations are realized  by sparse matrix-vector multiplications, and we use  a preconditioned 
conjugate gradient algorithm for solving the Poisson 
equation~(\ref{poisson}). 
The self-consistent solution to the set of 
Eq.~(\ref{eqstosolve}) is found by iteratively 
solving Eq.~(\ref{poisson}) and Eq.~(\ref{momentum}) until convergence is 
achieved. 
%
%

%
%
We begin the description of our MPS 
algorithm with a discussion of the encoding 
of discrete functions in MPS format. 
For this we assume that the number of grid 
points in each spatial dimension is 
$K=2^N$ for an integer $N$.
The binary representation 
$(\ldots)_2$ 
of a grid point  index $k^{\alpha}$ 
requires $\Ns$ bits,
\begin{equation}
 k^{\alpha}=\left(\sigma^{\alpha}_1,\sigma^{\alpha}_{2},\ldots,\sigma^{\alpha}_{\Ns}\right)_2\,,
\end{equation}
where $\sigma^{\alpha}_i\in\{0,1\}$, 
$\alpha\in\{x,y\}$, 
$i=1,\ldots,\Ns,$ and  $\sigma^{\alpha}_1$ and 
$\sigma^{\alpha}_{\Ns}$ are the most and least significant bits, respectively.
We approximate a discrete function $F$  
by an MPS of bond dimension $\chi$ and 
length $2N$, 
\begin{subequations}
\begin{align}
F(\mf{r}_k) \approx f(\mf{r}_k,  \chi) &= \underbrace{\mps^{\sigma_1^y}\mps^{\sigma_2^y}\cdots 
\mps^{\sigma_N^y}}_{\text{y-encoding}}
\;\underbrace{\mps^{\sigma_{1}^x}\cdots 
\mps^{\sigma_{N}^x}}_{\text{x-encoding}}\,,\\[0.2cm]
& = \mps^{\omega_1}\mps^{\omega_2}\cdots 
\mps^{\omega_{2N}}\,,
\label{mpsdef}
\end{align}
\label{mpsF}
\end{subequations}
where we introduced 
\begin{align}
 \omega_n=\left\{\begin{array}{ll}
                  \sigma_n^y, & 1\le n\le N,\\[0.1cm]
                  \sigma_{n-N}^x, & N< n \le 2N.
                 \end{array}
 \right.
\end{align}
The matrices $\mps^{\omega_n}$ have 
dimensions 
$\sm(n-1)\times\sm(n)$, where 
\begin{equation}
 \sm(n)=\min\left(2^n,2^{2N-n},\chi\right)
\end{equation}
are the internal bonds that are summed 
over in the product of matrices in 
Eq.~(\ref{mpsF}). These bonds are 
responsible for describing correlations 
between different length 
scales~\cite{gourianov:22,ng}.

The first $N$ matrices 
in Eq.~(\ref{mpsF}) encode the $y$-
components of $F$, and the 
remaining $N$ matrices account for the $x$-
components. 
Note that this encoding 
employs the  scale encoding 
introduced in~\cite{gourianov:22,ng} in
each spatial dimension separately. 
The encoding in Eq.~(\ref{mpsF}) 
thus corresponds to expanding 
the function $f$ as a 
sum of product functions, 
\begin{align}
 f(\mf{r}_k,\chi)=\sum\limits_{i=1}^{d(N)} 
 \mc{Y}_i\left(k^y\right) \mc{X}_i\left(k^x\right) \,,
\end{align}
where $\mc{Y}_i$ ($\mc{X}_i$) is a function 
of the $y$ index $k^y$ ($x$ index $k^x$) only. 
We find that this encoding is more efficient 
for the cavity geometry than the 
encoding in~\cite{gourianov:22} where 
combined scales of all spatial dimensions 
are considered.
Note that the encoding in Eq.~(\ref{mpsF}) can 
be straightforwardly generalized to the case 
where each spatial dimension is discretized by 
a different number of grid points. 
This is of interest for more complex 
geometries than the square box considered here. 

Next we describe how we emulate the DNS 
algorithm in tensor network format. 
The DNS algorithm can be broken down 
into the following elementary operations: 
(i) addition of flow fields, (ii) multiplication of flow fields, (iii)  the  algorithm for solving the Poisson equation,  and (iv) sparse matrix-vector operations. 
Sparse matrix-vector operations 
realize finite difference operations on 
the flow fields, as well as the boundary 
conditions for the vorticity in 
Eq.~(\ref{boundaryw}).        

Since all operations (i-iv) can be 
realized in MPS format, 
the MPS algorithm for solving 
Eq.~(\ref{eqstosolve}) can  be obtained 
by replacing each elementary operation 
in the DNS algorithm by its MPS counterpart.  MPSs can be added~\cite{schollwoeck:11} 
and multiplied~\cite{lubasch:18}, and 
a Poisson solver in MPS format has been reported in~\cite{oseledets:12}.  Matrix-vector operations are realized by contracting a matrix product 
operator (MPO) with an MPS~\cite{schollwoeck:11}, and 
all MPOs for realizing the finite 
difference operations and boundary 
conditions are provided in Appendix~\ref{mpos}. 
The numerical complexity of all these 
operations scales polynomially with 
the bond dimension $\chi$ of the MPS 
[for details see Appendix~\ref{algo}].  
It follows that the MPS realizations of 
operations (i-iv) can be numerically more 
efficient than their standard 
implementations for sufficiently small $
\chi$.

All variables (velocity components $u$ and $v$, 
streamfunction $\psi$ and vorticity $w$) are approximated 
by an MPS with bond dimension $\chi$. We allow $\chi$ 
to dynamically grow in order to keep the numerical 
complexity of our algorithm minimal. We achieve this 
by normalizing the MPSs representing $\psi$ and $w$ to unity, and by inspecting the singular values near 
the center 
of these MPSs. We increase the bond dimension if the 
smallest singular value exceeds a threshold $\epsilon$, 
which we set to $\epsilon=5\times10^{-8}$ throughout 
this work.  
This choice has been informed by numerical tests, 
ensuring  that all precision targets of the 
algorithms implementing the elementary operations 
are met with the smallest possible $\chi$. 

%
\begin{figure}[t!] 
\begin{center}
\includegraphics[width=\columnwidth]{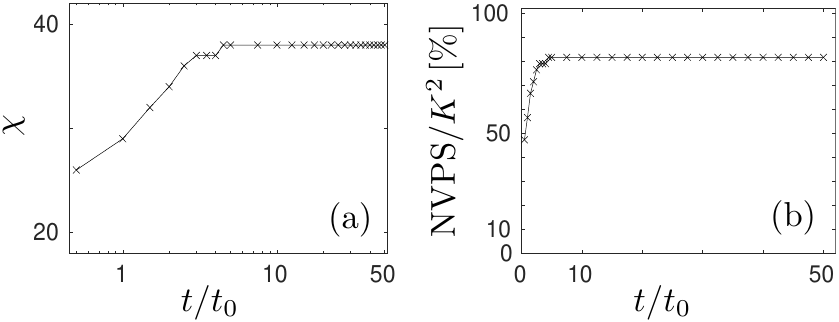}
\end{center} 
\caption{\label{fig2}
(a) Bond dimension $\chi$ versus time on a 
logarithmic scale and for the flow in 
Fig.~\ref{fig1}(b). 
(b) The ratio between the NVPS and the total 
number of grid points $K^2$ in percent and  as a function of time.
Solid lines are a guide to the eye. 
}
\end{figure}
%
\section{Results \label{results}}
In a first step we validate the  MPS algorithm against the 
tabulated results for the stationary 
state of the lid-driven cavity 
in~\cite{ghia:82}. 
We consider a $2^7\times 2^7$ grid ($N=7$) 
and  Reynolds number Re=1,000. The  
contours showing the velocity magnitude  according to the MPS algorithm and 
for $t/t_0=50$ are shown in 
Fig.~\ref{fig1}(b). We compare this to the 
data in~\cite{ghia:82} in Figs.~\ref{fig1}(b) 
and~(c). The velocity component $u$ along the 
vertical line $\mc{L}_v$ [see Fig.\ref{fig1}(a)] 
according the the MPS algorithm (black solid lines) agrees very well with the 
data in~\cite{ghia:82} (red dots). Similarly, 
we find that our MPS results for $v$ along the horizontal line $\mc{L}_h$ agree very well 
with~\cite{ghia:82} as shown in 
Fig.~\ref{fig1}(d). 
We note that our DNS algorithm is in excellent 
agreement with the MPS algorithm and with 
the reference data in~\cite{ghia:82}.

%
%
The MPS algorithm dynamically adapts 
the bond dimension of the MPS representing 
the flow fields. Initially the fluid it 
as rest, $u=v=0$. This constant velocity 
field is  an MPS with bond 
dimension $\chi=1$. However, we find through 
numerical experiments 
that the sudden quench induced  
by the moving lid requires a starting bond 
dimension of $\chi=26$. The subsequent 
evolution of $\chi$ with time for 
the flow in Fig.~\ref{fig1}(b) is shown in 
Fig.~\ref{fig2}(a). We find that $\chi$ 
approximately grows  logarithmically with time until 
$t/t_0 \approx 3$, and then it stays constant at $\chi=38$. 
At $t/t_0 \approx 3$, the  vortex created by the 
moving lid has expanded from the top right corner 
to the whole size of the cavity. While the vortex 
changes shape until the steady state is reached, 
the bond dimension stays constant in this regime. 

The bond dimension $\chi$ is directly 
related to the NVPS, which is shown 
in Fig.~\ref{fig2}(b) in relation to 
the total number of grid points $K^2$. 
Initially the NVPS are about 
$47\%$ of  $K^2$. For larger times, the NVPS slowly 
increases to  $82\%$ of $K^2$. 
%
%
It follows that the MPS format does not result 
in an efficient compression of the stationary state 
for Re=1,000.

\begin{figure}[t!] 
\begin{center}
\includegraphics[width=\columnwidth]{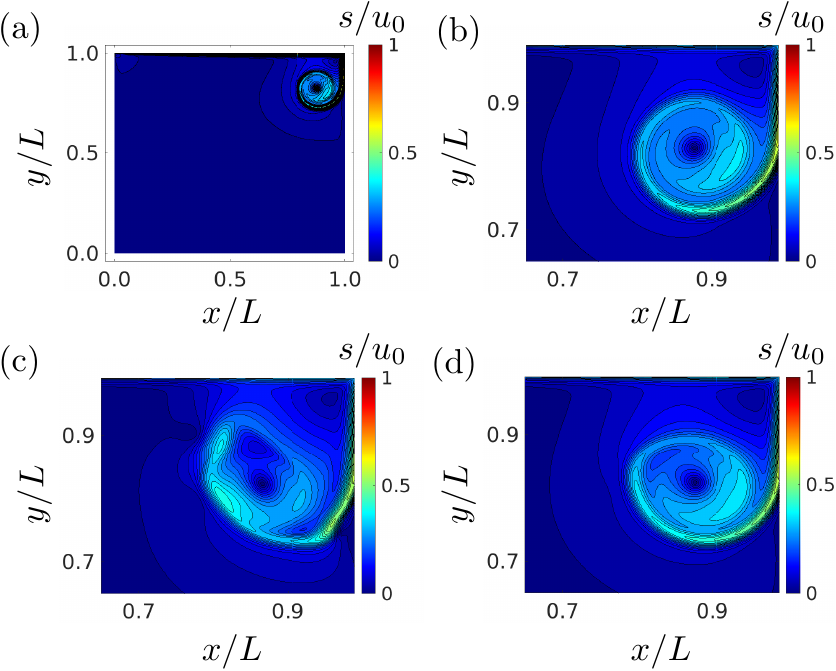}
\end{center} 
\caption{\label{fig3} 
(Color online) (a) Contour plot of the 
velocity magnitude $s=\sqrt{u^2+v^2}$ at $t/t_0=3$ 
for the flow configuration shown in 
Fig.~\ref{fig1}(a). The grid size is 
$K^2=2^{11}\times2^{11}$, Re=24,000 and 
results are obtained with the MPS algorithm.     
(b) Same as in (a) but focussing on the 
region where the initial vortex forms.
(c) Same as in (b) but for $K^2=2^{9}\times2^{9}$. 
(d) Same as in (b) but for $K^2=2^{10}\times2^{10}$. 
}
\end{figure}

The situation is completely different in a
transient regime at high Reynolds numbers. 
For this we consider a flow with  $\text{Re}=24,000$, 
and Fig.~\ref{fig3}(a) shows the  corresponding 
contours of the velocity magnitude on a $2^{11}\times2^{11}$ grid 
at $t/t_0=3$. A magnified view of the vortex 
forming in the top right corner is shown 
in Fig.~\ref{fig3}(b). 
It is well known that the lid-driven cavity only exhibits a truly 
stationary state for $\text{Re}\le 10,000$~\cite{verstappen:94}. For larger 
Reynolds numbers, the system becomes chaotic and develops random fluctuations that persist for all times. 
However, we find that at the short times 
considered here where turbulence has not 
formed yet, the system is still 
deterministic. All 
runs with the same initial conditions 
give the same result. We expect the  
onset of  turbulence and non-stationary 
fluctuations at much later times  
when the vortex has 
spread to the whole cavity. 
%
%

Next we investigate the required grid size 
to correctly represent this transient flow. 
For this we run the calculation for different 
grid sizes $2^{N}\times2^{N}$ with $N=8,9,10,11$ and 
$12$. The results for $N=9$ and $N=10$ are 
shown in Figs.~\ref{fig3}(c) and~(d),
respectively. By comparing it to 
the solution for $N=11$ in Fig.~\ref{fig3}(b), 
we find that the flow fields 
are underresolved on the $N=9,10$ grids. 
For $N=9$ [see Fig.~\ref{fig3}(c)], the vortex in the upper right 
corner is strongly deformed. The amount 
of deformation is much smaller but still 
 visible for the $N=10$ grid [see Fig.~\ref{fig3}(d)]. On 
the other hand, increasing the size to $N=12$ (not shown)
does not result in any significant changes 
compared with the results for $N=11$. 
We thus conclude that the $2^{11}\times2^{11}$ 
grid is sufficiently large for representing 
this flow.

The smallest length scale in a fully 
developed turbulent flow is the 
Kolmogorov microscale $\eta/L\approx \text{Re}^{-3/4}$~\cite{monin:71,monin:75}.
Although the flow investigated in Fig.~\ref{fig3} is not in the turbulent 
regime yet, the value of 
$\eta/L\approx 5.19\times10^{-4}$ for 
Re=24,000 is consistent with the 
grid point spacing $2^{-11}\approx 4.88\times 10^{-4}$ for the $2^{11}\times 2^{11}$ grid 
that resolves this flow. We conclude that 
the smallest scale according to Kolmogorov 
theory is excited even in the investigated 
regime where the flow is still laminar. 
It follows that $\eta/L$ gives a reasonable 
estimate for the required grid size. 
%

\begin{figure}[t!]  
\begin{center}
\includegraphics[width=\columnwidth]{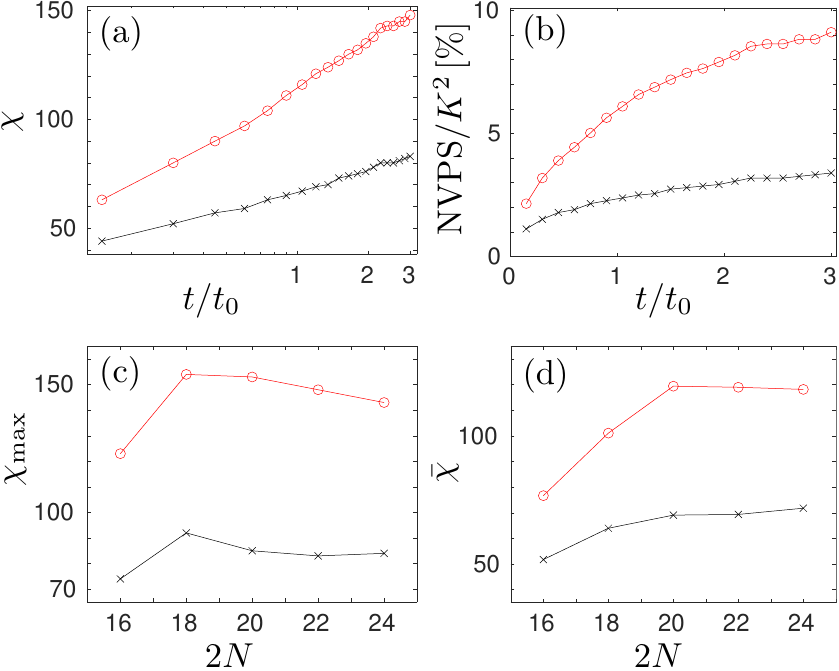}
\end{center} 
\caption{\label{fig4}
Analysis of the bond dimension as a function 
of time and grid size.  
Black crosses [red circles]
correspond to the flow in Fig.~\ref{fig3}(a) 
[Fig.~\ref{fig5}], and solid lines are 
a guide to the eye. 
(a) Bond dimension $\chi$ versus time on a 
logarithmic scale. 
(b) The ratio between the NVPS and the total 
number of grid points $K^2$ in percent and  as a function of time.  
(c) Bond dimension $\chi_{\text{max}}$ at $t/t_0=3$ as a 
function of  grid size.
(d) Temporally averaged bond dimension $\bar{\chi}$ as a 
function of  grid size.
%
%
}
\end{figure}

%
%
The variation of $\chi$ with 
time and for 
the flow in Fig.~\ref{fig3}(a) is shown by the 
black crosses in 
Fig.~\ref{fig4}(a). At $t=0$ 
we set $\chi=40$, and after a short initial 
phase [not shown in Fig.~\ref{fig4}(a)] 
we find that $\chi$ grows  
logarithmically in time. 
The corresponding NVPS in relation to 
the total number of grid points  is shown 
by the black crosses 
in Fig.~\ref{fig4}(b).   
At very short times, the NVPS are only about 
$1\%$ of  $K^2$, and thus the MPS format achieves a compression of $99\%$. For larger times, the NVPS slowly increases to  $3.4\%$ of $K^2$, 
corresponding to a compression of $96.6\%$. 
%
 
%
%
 
%
Next we investigate the dependence of 
the bond dimension on the grid size. 
We find that for all studied grids 
($N=8,9,10,11,12$), $\chi$ vs. time 
has the same qualitative behavior as 
shown in Fig.~\ref{fig4}(a) for $N=11$. 
For each of these curves, we calculate 
the maximal value $\chi_{\text{max}}$ at $t/t_0=3$ and 
the temporally averaged bond dimension 
$\bar{\chi}$. The results for $\chi_{\text{max}}$ and $\bar{\chi}$ 
are shown by black crosses in  Figs.~\ref{fig4}(a) and~(b), respectively.  
We find that  $\chi_{\text{max}}$ and $\bar{\chi}$ vary 
with $2N$ until the  grid is fine enough to represent
the flow. While $\bar{\chi}$ increases steadily with $2N$, 
$\chi_{\text{max}}$ first increases 
then decreases with $2N$.

\begin{figure}[t!]  
\begin{center}
\hspace*{0.5cm}
\vcenteredbox{
\normalsize{(a)}\\[0.6cm]
\begin{tabular}{c||c|c|c|c|}
 & $\mc{C}_t$ & $\mc{C}_r$ & $\mc{C}_b$ & $\mc{C}_l$\\
 \hline
 \hline
 $u$ & $u_0$ & 0 & $-u_0$ & 0 \\
 \hline
 $v$ & 0 & 0 & 0 & 0 \\
 \hline
 $\psi$ & 0 & 0 & 0 & 0 \\
 \hline 
\end{tabular}\\[1.4cm]}
\hspace*{\fill} 
\vcenteredbox{
\includegraphics[width=0.5\columnwidth]{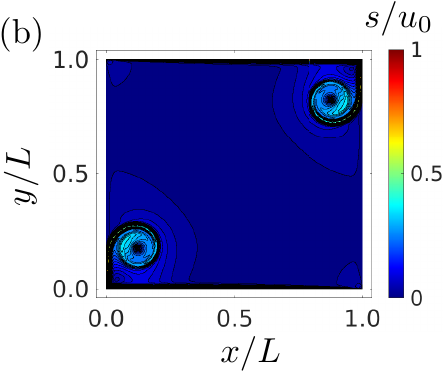}} 
\end{center} 
\caption{\label{fig5} 
(Color online) 
(a) Boundary conditions corresponding to the doubly-driven cavity 
where the upper [bottom] lid moves at constant velocity $u_0$ [$-u_0$]
in $x$-direction. 
(b) Contour plot of the 
velocity magnitude $s=\sqrt{u^2+v^2}$ at $t/t_0=3$ 
for the doubly-driven cavity on a $K^2=2^{11}\times2^{11}$ grid 
with  Re=24,000  and evaluated   with the 
MPS algorithm. 
}
\end{figure}

We now investigate how the  results for the 
bond dimension in the lid-driven cavity geometry change if we consider  
a doubly-driven cavity instead, see Fig.~\ref{fig5}. 
The upper lid continues to 
move at constant velocity $u_0$ in $x$-direction. In addition,  
the bottom lid moves at constant velocity $-u_0$ in $x$-direction. 
The corresponding contours of the velocity magnitude on a 
$2^{11}\times2^{11}$ grid and with Re=24,000 are shown in Fig.~\ref{fig5}(b).  We find that a second vortex forms in the 
bottom left corner of the cavity. The corresponding results 
for the bond dimension as a function of time and grid size are 
shown by the red circles in Fig.~\ref{fig4}. The qualitative 
behavior of all curves is similar to the lid-driven cavity, 
but the bond dimension for the doubly-driven cavity is larger 
than for the lid-driven cavity at each point in time. 
The NVPS in relation to the total number of grid points grows 
to about $9\%$ for the doubly-driven cavity, and hence the 
MPS format still achieves a compression of more than $90\%$. 

%
%
%
The results in Fig.~\ref{fig4} show that the bond dimension 
only grows logarithmically with simulation time, and that the 
MPS format achieves an efficient compression of the flow fields. 
The numerical complexity of the MPS algorithm depends on the 
bond dimension $\chi$ as detailed in Appendix~\ref{algo}. 
While the most costly operation is the multiplication of 
two MPSs, the algorithm spends the most time on solving 
the Poisson equation which scales as  
$2N\chi^3$~\cite{oseledets:12}. 
On the other hand, the DNS algorithm 
can be broken down into sparse matrix-vector 
multiplications scaling with the total number 
of grid points $K^2=2^{2N}$. 
The exponentially worse scaling of the DNS 
algorithm 
with respect to the number of grid points 
$K^2$ suggests that the MPS 
algorithm can give rise  to a computational 
advantage for sufficiently small values of $\chi$. 

To address this question we compare 
the runtimes of the MPS and DNS
algorithms. In 
order to achieve a fair comparison, we implemented 
the DNS and MPS algorithms in the same 
programming language (i.e., Matlab~\cite{MA}), and 
evaluated all runs on a single node of the 
ARC facility (Intel Xeon Platinum 8268 CPU @ 2.90GHz)~\cite{Andrews2015}. 
Furthermore, we ensure that the DNS and MPS 
algorithms solve Eq.~(\ref{eqstosolve}) with the same accuracy (see Appendix~\ref{algo}). 
We find that the MPS algorithm is 5.8 times faster 
than the DNS algorithm in the case of the lid-driven cavity. 
The speedup reduces to 3.3 for the 
doubly-driven cavity since the bond dimension 
is larger than for the lid-driven 
cavity at each time step, see Fig.~\ref{fig4}(a). 

A more comprehensive runtime comparison of the MPS 
and DNS algorithms at different Reynolds numbers 
is presented in Fig.~\ref{fig5}(a). 
The grid spacing for each Re is chosen such that it matches the 
corresponding microscale $\eta/L\approx \text{Re}^{-3/4}$. 
We show the ratio of the average times $T_{\text{DNS}}$ 
for completing a single iteration of the DNS algorithm 
and $T_{\text{MPS}}$ 
for completing a single iteration of the MPS algorithm. 
Since the MPS and DNS algorithms approximately require the 
same number of iterations, this ratio is also representative of 
the overall runtime ratio.
The MPS algorithm for the lid-driven and doubly-driven cavities 
runs faster than the DNS algorithm for $\text{Re}\ge9.5\times10^3$. 
For a given Reynolds number, the MPS algorithm for the doubly-driven 
cavity takes more time than in the case of the lid-driven cavity 
because the former requires larger bond dimensions, 
see Fig.~\ref{fig5}(b). For the largest Reynolds number,  the MPS algorithm approximately achieves 
a seventeen-fold [ten-fold] speedup compared with the DNS algorithm for the lid-driven [doubly-driven] cavity. 

\begin{figure}[t!]  
\begin{center}
\vcenteredbox{
\includegraphics[width=0.5\columnwidth]{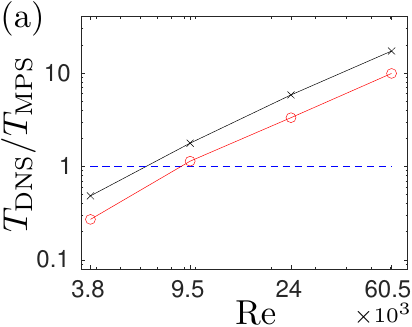}}  
\hspace*{\fill} 
%
\vcenteredbox{
\normalsize{(b)}\\[0.6cm]
\begin{tabular}{l||r|r|}
 & $\bar{\chi}(\text{LD})$ & 
 $\bar{\chi}(\text{DD})$ \\
 \hline
 \hline
 Re=$3.8$k & 46.3 & 71.8  \\
 \hline
 Re=$9.5$k & 55.5 & 91.7 \\
 \hline
 Re=$24$k & 69.3 & 119.0 \\
 \hline
 Re=$60.5$k & 112.7 & 152.7 \\
 \hline 
\end{tabular}\\[1.4cm]}
\end{center} 
\caption{\label{fig6} 
(Color online)  
(a)  Ratio of the average times $T_{\text{MPS}}$ 
for completing a single iteration of the MPS algorithm 
and $T_{\text{DNS}}$ 
for completing a single iteration of the DNS algorithm as a function
of Reynolds number Re. 
Black crosses [red circles]
correspond to the lid-driven [doubly-driven] cavity.
Averages are taken up to $t/t_0=3$. 
$T_{\text{DNS}}$  for 
Re=$24\text{k}$ (Re=$60.5\text{k}$) is only taken for 
$t/t_0\le1$ ($t/t_0\le0.1$) due to the large runtimes, and $T_{\text{MPS}}$
for the doubly-driven cavity and Re=$60.5\text{k}$ 
is evaluated for $t/t_0=2.1$.
The grid spacing 
for each Re is chosen such that it matches the 
corresponding microscale $\eta/L\approx \text{Re}^{-3/4}$. 
For data points above (below) the horizontal blue dashed line, the MPS (DNS) algorithm runs faster
than its DNS (MPS) counterpart. 
Solid lines are a guide to the eye.  
(b) Time-averaged bond dimensions $\bar{\chi}(\text{LD)}$ 
[$\bar{\chi}(\text{DD)}$] corresponding to the lid-driven [doubly-driven] cavity for different Reynolds numbers.
}
\end{figure}

The speedups shown in Fig.~\ref{fig6} 
can be qualitatively explained by noting that the DNS algorithm 
scales like $\text{Re}^{6/4}$, whereas the 
MPS algorithm scales as $\log\text{Re}$ for 
fixed bond dimension. However, the bond dimension 
grows with time and with Reynolds number, and 
therefore a general scaling of the runtime ratio 
with Reynolds number is difficult to obtain. 
At larger simulation times, 
the runtime advantage of the MPS algorithm may 
decrease or vanish if the required bond dimension 
becomes too large. 
The results in Fig.~\ref{fig6} nevertheless 
illustrate the 
tremendous potential of MPS for simulating 
transient flows.
%
%
\section{Summary and Discussion\label{summary}}
We have shown that dynamical solutions to the 
incompressible Navier-Stokes equations for 
the lid-driven and doubly-driven cavities can be obtained via a 
tensor network algorithm. Our work extends the results 
in~\cite{gourianov:22} by showing that 
the tensor network approach is not restricted to 
periodic boundary conditions but works equally 
well for problems with fixed boundary conditions. We achieve 
this by decomposing a DNS algorithm  
based on MacCormack's 
method~\cite{MacCormack:69,rider:05,anderson:cfd} 
into four elementary operations of addition, 
multiplication, matrix-vector multiplication and solving 
the Poisson equation. These four operations can be 
implemented in MPS format and the resulting MPS algorithm automatically builds 
a ROM characterized by a bond dimension $\chi$.  
Note that this  ROM becomes exact 
with sufficiently large  bond dimension, which  
distinguishes it from 
data-driven ROMs~\cite{ddse:19,kutz:dmd,ramezanian:21} for CFD 
which lack this guarantee of success.

It is important to note that our 
approach also applies to other CFD methodologies and flow geometries. 
For example, the streamfunction-vorticity formulation chosen 
in this work can be replaced with continuity and 
momentum equations expressed in terms of velocities and 
pressure~\cite{soh:88}. Re-writing this algorithm in 
terms of tensor network operations follows the same route as 
presented here. 
We run the MPS algorithm on a uniform grid and find 
that it automatically allocates resources only to those regions in space 
where they are needed.  No a priory knowledge of the flow is required. 
For example, the NVPS required by MPS to 
describe the  transient regime at large Reynolds number 
is at most 3\% of the total number of gridpoints. 
This very efficient MPS representation of the flow 
occurs because the vortex only
occupies  a small region in space. 
Very little resources are needed to represent the flow in 
the large area where the fluid is nearly at rest, 
see Fig.~\ref{fig3}(a). 
Adding the second vortex in the case of the doubly-driven 
cavity increases the NVPS to 9\%. 

A related finding 
is that the bond dimension of the MPSs 
representing the flow fields is approximately 
constant if the grid is fine enough to 
represent the flow. 
This feature is 
related to the known fact that 
polynomials and Fourier series 
have efficient MPS representations where the bond dimension is 
independent of the grid size~\cite{khoromskij:11,oseledets:13}. 
This behavior is also akin to one-dimensional quantum 
systems obeying an area law~\cite{Eisert2010}.

We find that the bond dimension of the MPSs representing the transient flows investigated 
in this work  grows logarithmically in time. This 
slow increase can translate into a runtime advantage of the 
MPS  vs.  DNS algorithms if the bond dimension 
of the initial flow fields is sufficiently small. 
We find that the MPS algorithm can be significantly 
faster than the DNS algorithm for 
simulation times of several units of $t_0 = L/u_0$, 
i.e., the time it takes the lid to traverse the 
length $L$ of the cavity. 
In general, our analysis shows that the MPS 
algorithm will outperform the DNS algorithm 
at large Reynolds numbers, provided that 
the required  bond dimension 
is sufficiently small. We anticipate that 
the maximal bond dimension allowing for 
a speedup depends on the used hardware and 
software implementation of the algorithm, 
which is subject to further study.

Several avenues for further research emerge from 
here. First, the transient flow example studied in this work  
may also be efficiently described  with adaptive 
mesh refinement~\cite{berger:84,berger:89}. In this approach, 
the grid spacing is dynamically varied in space 
at the cost of detecting the areas requiring 
high-resolution grids. 
It would be interesting to directly compare the performance 
of these two methods for different flow types, and to 
establish the differences and similarities between them.

Second,   MPS algorithms for CFD may benefit from 
modern hardware architectures optimized for 
tensor operations~\cite{ganahl:22}. 
This opens up the exciting prospect of 
developing tensor network algorithms for 
technical flows that outperform state-of-the-art 
CFD algorithms.

Finally,  CFD 
algorithms in tensor network format represent 
a first step towards solving the Navier-Stokes 
equations on a quantum computer~\cite{jaksch:22,griffin:19,lubasch:20}. 
Quantum CFD~\cite{gourianov:22,fukagata:22} 
promises to enable DNS for analyzing and optimizing technical flows, which would represent a revolutionary improvement of the 
state-of-the-art~\cite{jaksch:22}.  
Creating and benchmarking quantum CFD algorithms for wall-bounded flows 
by porting tensor network algorithms 
to quantum hardware is thus an exciting prospect for future research.

\begin{acknowledgments} 
M.K. acknowledges support by EPSRC Programme Grant DesOEQ (EP/P009565/1) 
and thanks D. Peshkin, P. Moinier, H. Babaee, T. Rung,  S. Bengoechea, P. Over, and L. Anderson for discussions.
D.J. acknowledges support by the European Union’s Horizon Programme (HORIZON-CL4-2021-DIGITALEMERGING-02-10) Grant Agreement 101080085 QCFD, by EPSRC Programme Grant
DesOEQ (EP/P009565/1), from AFOSR grant FA8655-22-1-7027 and by the Excellence
Cluster ‘The Hamburg Centre for Ultrafast Imaging—Structure, Dynamics and Control of Matter at the Atomic Scale’
of the Deutsche Forschungsgemeinschaft.
\end{acknowledgments}
%
%
\appendix

\section{DNS algorithm \label{dns}}
The DNS algorithm for solving 
Eq.~(\ref{eqstosolve}) can be broken down 
into four steps for advancing the solutions 
for $w$, $\psi$, $u$ and $v$ from time $t$ 
to $t+\Delta t$. In the following we describe 
each of these steps:

(i) Starting with the  streamfunction $\psi^t$
at time $t$, we calculate 
the  velocity components $u$ and $v$ according 
to Eq.~(\ref{velocities}). For this 
we employ a second-order accurate central 
difference approximation of the first  
derivatives in $x-$ and $y$-direction, 
\begin{subequations}
\begin{align}
[\partial_x \psi]_{p,q} &= 
\frac{1}{2h}(\psi_{p+1,q}-\psi_{p-1,q})\,, 
 \label{dx}\\
[\partial_y\psi]_{p,q} & = 
\frac{1}{2h}(\psi_{p,q+1}-\psi_{p,q-1})\,, 
\label{dy}
\end{align}
\label{dxy}
\end{subequations}
(ii) The vorticity is propagated in time by 
an explicit, second-order accurate 
MacCormack scheme~\cite{MacCormack:69,rider:05,anderson:cfd}.
To this end we write Eq.~(\ref{momentum}) as 
\begin{align}
\partial_t w & = \partial_x F +\partial_y G\,, 
\label{wfg}
\end{align}
where 
\begin{subequations}
\begin{align}
 F & = -u w +\nu(\partial_x w)\,, \label{f} \\
 G & = -v w +\nu(\partial_y w)\,.
 \label{g}
\end{align}
\label{FG}
\end{subequations}
MacCormack's algorithm advances $w^t$ to 
$w^{t+\Delta t}$ in a two-step predictor-corrector 
procedure:\\[0.3cm]
$\bullet$ Predictor step: \\[0.3cm]
In order to evaluate $F$ and $G$, the derivatives $\partial_x w$ and $\partial_y w$ 
in Eq.~(\ref{FG}) are approximated by first-order 
accurate backward differences 
 $\delta_x^{\text{bwd}}$ and $\delta_x^{\text{bwd}}$, respectively,
\begin{subequations}
\begin{align}
[\delta_x^{\text{bwd}}w]_{p,q} & = 
\frac{w_{p,q}-w_{p-1,q}}{h} \,,\\
[\delta_y^{\text{bwd}}w]_{p,q} & = 
\frac{w_{p,q}-w_{p,q-1}}{h}\,.
\end{align}
\label{delbwd}
\end{subequations}
The predicted solution $\bar{w}^{t+\Delta t}$ 
(indicated by an overbar) is obtained by 
a first-order accurate forward discretisation 
of the spatial derivatives in Eq.~(\ref{wfg}), 
\begin{align}
   \bar{w}_{p,q}^{t+\Delta t} & = w_{p,q}^t + 
 \left(\frac{F_{p+1,q}^t - F_{p,q}^t}{h} 
   + \frac{G_{p,q+1}^t -
   G_{p,q}^t}{h}\right)\Delta t\,.
  \label{predictor}
 \end{align}
Evaluating Eq.~(\ref{delbwd}) on 
the inner grid with  
$p,q \in \{0,\ldots,K-1\}$ 
requires the boundary values of $w$ for $w_{-1,q}$ 
and $w_{p,-1}$ in Eq.~(\ref{boundaryw}).
In addition, Eq.~(\ref{predictor}) for 
$p=K-1$ requires $[\delta_x^{\text{bwd}}w]_{K,q}$, and for $q=K-1$ we need 
$[\delta_y^{\text{bwd}}w]_{p,K}$. These 
values can be obtained with the help 
the boundary values $w_{K,q}$ and $w_{p,K}$, 
respectively. \\[0.3cm]
$\bullet$ Corrector step: \\[0.3cm]
The 
derivatives $\partial_x w$ and $\partial_y w$ in Eq.~(\ref{FG}) 
are now approximated by first-order accurate 
forward differences 
 $\delta_x^{\text{fwd}}$ and $\delta_x^{\text{fwd}}$, respectively,
\begin{subequations}
\begin{align}
[\delta_x^{\text{fwd}}w]_{p,q} & = 
\frac{w_{p+1,q}-w_{p,q}}{h}\,, \\
[\delta_y^{\text{fwd}}w]_{p,q} & = 
\frac{w_{p,q+1}-w_{p,q}}{h}\,.
\end{align}
\label{delfwd}
\end{subequations}
We update the functions $F$ and $G$ with the 
predicted solution $\bar{w}^{t+\Delta t}$ and 
obtain $\bar{F}^{t+\Delta t}$ and 
$\bar{G}^{t+\Delta t}$. 
The solution for 
the vorticity $w^{t+\Delta t}$ at $t+\Delta t$ 
is then obtained by approximating the 
spatial derivatives in Eq.~(\ref{momentum}) 
by first-order accurate backward differences, 
\begin{align}
   w_{p,q}^{t+\Delta t} & = \frac{1}{2} 
   \left(w_{p,q}^t +  \bar{w}_{p,q}^{t+\Delta t}\right) \notag \\ 
 & + \frac{1}{2}\left(\frac{\bar{F}_{p,q}^{t+\Delta t} - \bar{F}_{p-1,q}^{t+\Delta t}}{h} 
   + \frac{\bar{G}_{p,q}^{t+\Delta t} -
   \bar{G}_{p,q-1}^{t+\Delta t}}{h}\right)\Delta t\,.
  \label{corrector}
 \end{align}
With the help of the boundary values for 
$w$ in Eq.~(\ref{boundaryw}), Eq.~(\ref{corrector}) can be evaluated on 
every point of the inner grid 
with $p,q \in \{0,\ldots,K-1\}$. 
Although the forward- and backward differences 
in Eqs.~(\ref{predictor})-(\ref{corrector}) 
are only first-order accurate in $h$, 
the resulting expression for 
$w_{p,q}^{t+\Delta t}$ in 
Eq.~(\ref{corrector}) is second-order accurate~\cite{MacCormack:69,rider:05,anderson:cfd}. 

(iii) The vorticity $w^{t+\Delta t}$ is used 
to find the streamfunction $\psi^{t+\Delta t}$ 
at time $t+\Delta t$ by solving  
Eq.~(\ref{poisson}) with the boundary conditions 
in Tab.~\ref{tabB} and a second-order 
accurate discretisation of the Laplace operator, 
\begin{align}
 [\Delta\psi]_{p,q}=\frac{\psi_{p+1,q}+\psi_{p-1,q}
 +\psi_{p,q+1}+\psi_{p,q-1}-4 \psi_{p,q}}{h^2}\,.
\end{align}
(iv) The set of equations~(\ref{eqstosolve}) 
are coupled because the updated streamfunction 
$\psi^{t+\Delta t}$ gives rise to new 
velocity components $u^{t+\Delta t}$ and 
$v^{t+\Delta t}$ via Eq.~(\ref{velocities}). We repeat steps (i)-(iii) until 
a self-consistent solution to Eq.~(\ref{eqstosolve}) 
has been found. This results in updated 
functions $\psi^{t+\Delta t}$, $w^{t+\Delta t}$, 
$u^{t+\Delta t}$ and $u^{t+\Delta t}$ and completes 
the time step from $t$ to $t+\Delta t$. We 
repeat steps (i)-(iv) until the final time 
is reached. 
\section{MPS Algorithms \label{algo}}
\begin{table}[t!]
 
\centering

\begin{tabularx}{\columnwidth}{p{2cm}||X|P{1.2cm}|}
 Operation & Algorithm  & Scaling \\
 \hline
 \hline
\vspace*{-0.1cm}

Addition
& Variational addition of MPS (see  Sec. 4.5 in~\cite{schollwoeck:11}). & 
\vspace*{-0.1cm} 
$\chi^3$ \\
 \hline
 \vspace*{0.1cm}
 Multiplication & Multiplication algorithm 
 in~\cite{lubasch:18} combined with variational compression~\cite{schollwoeck:11} of the product MPS. & 
\vspace*{0.1cm}
 $\chi^4$ \\
 \hline
 \vspace*{-0.1cm}
 Poisson solver & MPS algorithm for 
 solving the Poisson equation in~\cite{oseledets:12}. & 
\vspace*{-0.1cm}
 $\chi^3$ \\
 \hline 
 \vspace*{0.35cm}
Matrix-vector multiplication &  MPO-MPS contraction combined with variational 
compression (see  
  Sec. 5 in~\cite{schollwoeck:11}). 
  For the system considered here, the 
  MPO bond dimension $D\le6$ and thus $D\ll\chi$. 
  & 
\vspace*{0.4cm}
 $D\chi^3$ \\
 \hline
\end{tabularx}
\caption{\label{tabA} Overview of the 
algorithms for realising the building 
blocks of the DNS algorithm in MPS format. 
The last column indicates the scaling 
of the operation 
with the bond dimension of the MPSs and MPOs.}
\end{table}
Table~\ref{tabA} outlines the MPS algorithms for 
realising the required elementary operations as well as 
their scaling with the bond dimension $\chi$. 
All these algorithms have in common that they are 
variational in nature. The desired MPS for representing 
the target, i.e., the sum or product of MPSs or the solution to the Poisson equation, is found by minimising a 
cost function. These cost functions are quadratic in 
the variables and hence efficient and reliable methods 
for finding optimal solutions exist. We employ single-site  
DMRG-like~\cite{schollwoeck:11} sweeps where each tensor 
in the MPS is sequentially optimised until overall 
convergence has been achieved. 

In order to make the results of the MPS algorithm 
comparable to the DNS results, we impose the same 
accuracy goal for solving the Poisson equation 
and the same convergence criterion for 
solving Eq.~(\ref{eqstosolve}) in both 
algorithms. 
\section{MPOs for finite difference operations 
 \label{mpos}}
%
%
Here we show how the required finite difference
operations can be created 
in the MPO-MPS formalism. We denote an MPO by $Q$ and 
its contraction with an MPS $f$ as $Qf$. 
A generic MPO with bond dimension $D$ 
can be written as~\cite{crosswhite:08} 
\begin{align}
 Q & = 
\lb \md[1]\cdots \md[N] \md[N+1]\cdots \md[2N] \rb\,,
 \label{mpo}
\end{align}
where $\lb$ is a $1\times D$ row vector, $\rb$ is a 
$D\times 1$ column vector,  
and   $\md[k]$ with $k\in\{1,\ldots,2N\}$ are 
$D\times D$ matrices whose 
matrix elements are $2\times2$ 
matrices. Any $2\times2$ matrix  can be expanded in terms of 
the following four operators, \\
\begin{subequations}
\begin{align}
 \sigma_{01} & =\left(
 \begin{array}{cc}
  0 & 1 \\
  0 & 0
 \end{array}
\right)\,, \\
 \sigma_{10} & =\left(
 \begin{array}{cc}
  0 & 0 \\
  1 & 0
 \end{array}
\right)\,, \\
\sigma_{00} & =\left(
 \begin{array}{cc}
  1 & 0 \\
  0 & 0
 \end{array}
\right)\,, \\
 \sigma_{11} & =\left(
 \begin{array}{cc}
  0 & 0 \\
  0 & 1
 \end{array}
\right)\,.
 \end{align}
 \end{subequations}
 For convenience, we also introduce the identity 
 matrix 
 \begin{align}
 \mathds{1} & =\left(
 \begin{array}{cc}
  1 & 0 \\
  0 & 1
 \end{array}
\right)\,.
\end{align}
When multiplying the matrices $\md[k]$ in 
Eq.~(\ref{mpo}), 
we take the outer product of the 
matrix-valued matrix elements. In order 
to illustrate this notation, we consider 
the following example for $N=1$,
\begin{align}
  \lb & = \left( 1, 0\right)\,, \\
 \md[k] & = \left(
 \begin{array}{cc}
   \mathds{1} & \sigma_{01}  \\
   0 & \sigma_{10}
 \end{array}
\right)\,, \quad 1\le k\le 2 \\
 \rb & = \left( 1, 1\right)^t\,.
\end{align}
The corresponding MPO is 
\begin{align}
 Q & = \left( 1, 0\right) 
 \left(\begin{array}{cc}
        \mathds{1}\otimes\mathds{1} & 
        \mathds{1}\otimes\sigma_{01} 
        + \sigma_{01}\otimes\sigma_{10} \\
        0 & \sigma_{10}\otimes\sigma_{10}
        \end{array}
\right) \left(\begin{array}{c}
               1 \\
               1
              \end{array}
\right) \\
 & = \mathds{1}\otimes\mathds{1}+\mathds{1}\otimes\sigma_{01}+\sigma_{01}\otimes\sigma_{10}\, ,
\end{align}
where $\otimes$ denotes the outer product.

The MPO representation of the first-order accurate 
forward-backward differences are described in 
Sec.~\ref{fwdbwd}, and Secs.~\ref{laplace} 
and~\ref{mpocentral} provide the MPOs 
for the Laplace operator and the  central differences, respectively.
\subsection{Forward-backward differences \label{fwdbwd}}
We provide generic expressions for the MPOs 
facilitating forward- and backward differences 
in Sec.~\ref{zerob}. These expressions 
are valid if  the boundary values 
of the function to be differentiated are zero 
everywhere. 
Specific expressions are required 
for functions with non-zero boundary values. 
In the algorithm described in Sec.~\ref{dns}, 
boundary values are required for 
calculating finite-difference approximations 
of the first and second spatial derivatives of $w$. These expressions are given in 
Appendices~\ref{predspec} and~\ref{corrspec} 
for the predictor and corrector steps, 
respectively.
\subsubsection{Generic expressions\label{zerob}}
$\bullet$ Forward-differencing in $x$-direction:
\begin{subequations}
\begin{align}
 [Q_x^{\text{fwd}}f]_{p,q}=\frac{f_{p+1,q}-f_{p,q}}{h}\,,
\end{align}
with
\begin{align}
 \lb & = \left( 1, 0\right)/h\,, \\
 \md[k] & = \mathds{1}\,, \quad 1\le k\le N\,, \\
 \md[k] & = \left(
 \begin{array}{cc}
   \mathds{1} & \sigma_{01}  \\
   0 & \sigma_{10}
\end{array}
\right)\,, \quad N< k \le 2N\,. \\
 \rb & = \left( -1, 1\right)^t\,. 
\end{align}
\label{fwdx}
\end{subequations}
$\bullet$ Backward-differencing in $x$-direction:
\begin{subequations}
\begin{align}
 [Q_x^{\text{bwd}}f]_{p,q}=\frac{f_{p,q}-f_{p-1,q}}{h}\,,
\end{align}
with
\begin{align}
 \lb & = \left( 1, 0\right)/h\,, \\
 \md[k] & = \mathds{1}\,, \quad 1\le k\le N\,, \\
 \md[k] & = \left(
 \begin{array}{cc}
   \mathds{1} & \sigma_{10}  \\
   0 & \sigma_{01}
\end{array}
\right)\,, \quad N< k \le 2N\,, \\
 \rb & = \left( -1, 1\right)^t\,. 
\end{align}
\label{bwdx}
\end{subequations}
$\bullet$ Forward-differencing in $y$-direction:
\begin{subequations}
\begin{align}
 [Q_y^{\text{fwd}}f]_{p,q}=\frac{f_{p,q+1}-f_{p,q}}{h}\,,
\end{align}
with
\begin{align}
 \lb & = \left( 1, 0\right)/h\,, \\
 \md[k] & = \left(
 \begin{array}{cc}
   \mathds{1} & \sigma_{01}  \\
   0 & \sigma_{10}
\end{array}
\right)\,, \quad 1\le k\le N\,, \\
 \md[k] & = \mathds{1}\,, \quad N< k \le 2N\,. \\
 \rb & = \left( -1, 1\right)^t\,. 
\end{align}
\label{fwdy}
\end{subequations}
$\bullet$ Backward-differencing in $y$-direction:
\begin{subequations}
\begin{align}
 [Q_y^{\text{bwd}}f]_{p,q}=\frac{f_{p,q}-f_{p,q-1}}{h}\,,
\end{align}
with
\begin{align}
 \lb & = \left( 1, 0\right)/h\,, \\
 \md[k] & = \left(
 \begin{array}{cc}
   \mathds{1} & \sigma_{10}  \\
   0 & \sigma_{01}
\end{array}
\right)\,, \quad 1\le k\le N\,, \\
 \md[k] & = \mathds{1}\,, \quad N< k \le 2N\,, \\
 \rb & = \left( -1, 1\right)^t\,. 
\end{align}
\label{bwdy}
\end{subequations}
\subsubsection{Finite differences of $w$ - 
predictor step\label{predspec}}
Here we provide the MPOs required for 
evaluating the predictor step in Eq.~(\ref{predictor}). \\[0.2cm]
$\bullet$ Backward-difference of $w$ in $x$-direction:
\begin{align}
 \delta_x^{\text{bwd}} w \approx Q_x^{\text{bwd}}f_w 
 -\frac{1}{h}Q_{\mc{C}_l} f_{\psi}=f_{w_x}^{\text{bwd}}\,,
 \label{wxbwd}
\end{align}
where $Q_x^{\text{bwd}}$ is given in 
Eq.~(\ref{bwdx}), and $f_{w}$ and $f_{\psi}$
are the MPSs representing $w$ and $\psi$, 
respectively. 
The MPO $Q_{\mc{C}_l}$ creates the boundary values for $w$ at $\mc{C}_l$, 
\begin{subequations}
\begin{align}
 [Q_{\mc{C}_l}f]_{p,q} & = \frac{1}{h^2}\left(- 4f_{0,q} + \frac{1}{2}f_{1,q} \right)\delta_{p,0}\,,
\end{align}
with
\begin{align}
 \lb & = 1/h^2\,, \\
 \md[k] & = \mathds{1}\,, \quad 1\le k \le N\,, \\
 \md[k] & = \sigma_{00}\,, \quad N< k\le 2N-1\,, \\
 \md[2N] & = -4\sigma_{00} +\sigma_{01}/2\,, \\
 \rb & = 1\,. 
\end{align}
\label{mpocl}
\end{subequations}
In Eq.~(\ref{wxbwd}), $f_{w_x}^{\text{bwd}}$ 
is the MPS representing 
$\delta_x^{\text{bwd}}w$. \\[0.2cm]
$\bullet$ Forward-backward-difference of $w$ in $x$-direction:
\begin{align}
 \delta_x^{\text{fwd}}(\delta_x^{\text{bwd}}w) 
 \approx &  Q_x^{\text{fwd}}f_{w_x}^{
 \text{bwd}}
+\frac{1}{h} \left[\frac{1}{h}(
 Q_{\mc{C}_r}f_{\psi}  - Q_{r}f_w)\right]\,,
 \label{wxx}
\end{align}
where $Q_{\mc{C}_r}$  is defined as 
\begin{subequations}
\begin{align}
 [Q_{\mc{C}_r} f]_{p,q} = \frac{1}{h^2}\left(- 4 f_{K-1,q} + \frac{1}{2}f_{K-2,q} \right) \delta_{p,0}\,,
\end{align}
with
\begin{align}
 \lb & = 1/h^2\,, \\
 \md[k] & = \mathds{1}\,, \quad 1\le k \le N\,, \\
 \md[k] & = \sigma_{11}\,, \quad N< k\le 2N-1\,, \\
 \md[2N] & = -4\sigma_{11} +\sigma_{10}/2\,, \\
 \rb & = 1\,. 
\end{align} 
\label{mpocr}
\end{subequations}
The MPO $Q_{r}$ in 
Eq.~(\ref{wxx}) extracts the values of 
a function on the line $k^x = K-1$,
\begin{align}
 [Q_{r}f]_{p,q}=f_{p,q}\delta_{K-1,q}\,,
\end{align}
with
\begin{subequations}
\begin{align}
 \lb & = 1\,, \\
 \md[k] & = \mathds{1}\,, \quad 1\le k \le N\,, \\
 \md[k] & = \sigma_{11}\,, \quad N< k\le 2N\,, \\
\rb & = 1\,. 
\end{align}
\end{subequations}
$\bullet$ Backward-difference of $w$ in $y$-direction:
\begin{align}
\delta_y^{\text{bwd}}w \approx Q_y^{\text{bwd}}f_w 
 -\frac{1}{h}Q_{\mc{C}_b} f_{\psi}= f_{w_y}^{
 \text{bwd}}\,,
 \label{wybwd}
\end{align}
where $Q_y^{\text{bwd}}$ is given in 
Eq.~(\ref{bwdy}), and $f_{w}$ and 
$f_{\psi}$ are the MPSs 
representing $w$ and $\psi$, 
respectively. 
The MPO $Q_{\mc{C}_b}$ creates the boundary 
values for $w$ at $\mc{C}_b$ and is defined as 
\begin{subequations}
\begin{align}
 [Q_{\mc{C}_b}f]_{p,q} & = \frac{1}{h^2}\left(- 4f_{p,0} + \frac{1}{2}f_{p,1} \right)\delta_{q,0} \,,
\end{align}
with
\begin{align}
 \lb & = 1/h^2\,, \\
 \md[k] & = \sigma_{00}\,, \quad 1\le k\le N-1\,, \\
 \md[N] & = -4\sigma_{00} +\sigma_{01}/2\,, \\
 \md[k] & = \mathds{1}\,, \quad N< k \le 2N\,, \\
 \rb & = 1\,. 
\end{align}
\label{mpocb}
\end{subequations}
In Eq.~(\ref{wybwd}), $f_{w_y}^{\text{bwd}}$ 
is the MPS representing 
$\delta_y^{\text{bwd}}w$. \\[0.2cm]
$\bullet$ Forward-backward-difference of $w$ in $y$-direction:
\begin{align}
 \delta_y^{\text{fwd}}(\delta_y^{\text{bwd}}w) 
 \approx &  Q_y^{\text{fwd}}f_{w_y}^{
 \text{bwd}}
 \notag \\
 & +\frac{1}{h} \left[\frac{1}{h}(
 Q_{\mc{C}_t}f_{\psi} + f_{u_0} - Q_{t}f_w)\right]
 \,, \label{wyy}
\end{align}
where $Q_{\mc{C}_t}$  is defined as 
\begin{subequations}
\begin{align}
 [Q_{\mc{C}_t}f]_{p,q} & = \frac{1}{h^2}\left(- 4f_{p,K-1} + \frac{1}{2}f_{p,K-2} \right)\delta_{q,K-1}\,,
 \end{align}
with 
\begin{align}
 \lb & = 1/h^2\,, \\
 \md[k] & = \sigma_{11}\,, \quad 1\le k\le N-1\,, \\
 \md[N] & = -4\sigma_{11} +\sigma_{10}/2\,, \\
 \md[k] & = \mathds{1}\,, \quad N< k \le 2N\,, \\
 \rb & = 1\,.
\end{align}
\label{mpoct}
\end{subequations}
The MPS  $f_{u_0}$ of bond dimension 1 
accounts for the $u_0$ term in the 
boundary condition~(\ref{uterm}). The 
matrices in the generic MPS definition~(\ref{mpsdef}) corresponding 
to $f_{u_0}$ are given by 
\begin{align}
M^{\omega_k}&=\left\{ 
\begin{array}{c}
 -3\frac{u_0}{h} \delta_{\omega_1, 1}
\,, \quad k=1 \,, \\[0.1cm]
\delta_{\omega_k, 1}\,,\quad 2\le k\le N\,,\\[0.1cm]
  1\,,\quad N<k\le 2N\,.
\end{array}
\right.
\label{mpsu}
\end{align}
Finally, the MPO $Q_{t}$ in 
Eq.~(\ref{wyy}) extracts the values of 
a function on the line $k^y = K-1$,
\begin{subequations}
\begin{align}
 [Q_{t}f]_{p,q}=f_{p,q}\delta_{q,K-1}\,,
\end{align}
with
\begin{align}
 \lb & = 1\,, \\
 \md[k] & = \sigma_{11}\,, \quad 1\le k \le N\,, \\
 \md[k] & = \mathds{1}\,, \quad N< k\le 2N\,, \\
 \rb & = 1\,. 
\end{align}
\end{subequations}
\subsubsection{Finite differences of $w$ - 
corrector step\label{corrspec}}
Here we provide the MPOs required for 
evaluating the corrector step in Eq.~(\ref{corrector}). \\[0.2cm]
$\bullet$ Forward-difference of $w$ in $x$-direction:
\begin{align}
 \delta_x^{\text{fwd}} w \approx Q_x^{\text{fwd}}f_w 
 + \frac{1}{h}Q_{\mc{C}_r} f_{\psi}=f_{w_x}^{\text{fwd}}\,,
 \label{wxfwd}
\end{align}
where $Q_x^{\text{fwd}}$ is given in 
Eq.~(\ref{fwdx}), and $f_{w}$ and $f_{\psi}$
are the MPSs representing $w$ and $\psi$, 
respectively. 
The MPO $Q_{\mc{C}_r}$ creates the boundary values for $w$ at $\mc{C}_r$ and is 
defined in Eq.~(\ref{mpocr}). 
In Eq.~(\ref{wxfwd}), $f_{w_x}^{\text{fwd}}$ 
is the MPS representing 
$\delta_x^{\text{fwd}}w$. \\[0.2cm]
$\bullet$ Backward-forward-difference of $w$ in $x$-direction:
\begin{align}
 \delta_x^{\text{bwd}}(\delta_x^{\text{fwd}}w) 
 \approx &  Q_x^{\text{bwd}}f_{w_x}^{
 \text{fwd}}
-\frac{1}{h} \left[\frac{1}{h}(
  Q_{l}f_w - Q_{\mc{C}_l}f_{\psi}  )\right]\,,
 \label{wxxc}
\end{align}
where $Q_{\mc{C}_l}$  is defined 
in Eq.~(\ref{mpocl}) and 
$Q_{l}$ is given by 
\begin{subequations}
\begin{align}
 [Q_{l}f]_{p,q}=f_{p,q}\delta_{0,q}\,,
\end{align}
with
\begin{align}
 \lb & = 1\,, \\
 \md[k] & = \mathds{1}\,, \quad 1\le k \le N\,, \\
 \md[k] & = \sigma_{00}\,, \quad N< k\le 2N\,, \\
\rb & = 1\,. 
\end{align}
\end{subequations}
$\bullet$ Forward-difference of $w$ in $y$-direction:
\begin{align}
\delta_y^{\text{fwd}}w \approx Q_y^{\text{fwd}}f_w 
 +\frac{1}{h}\left(
 Q_{\mc{C}_t} f_{\psi}+ f_{u_0}\right)
 = f_{w_y}^{\text{fwd}}\,,
 \label{wyfwd}
\end{align}
where $Q_y^{\text{fwd}}$ is given in 
Eq.~(\ref{fwdy}) and the MPS $f_{u_0}$ is 
defined in Eq.~(\ref{mpsu}). 
The MPO $Q_{\mc{C}_t}$  is 
defined in Eq.~(\ref{mpoct}). 
In Eq.~(\ref{wyfwd}), $f_{w_y}^{\text{fwd}}$ 
is the MPS representing 
$\delta_y^{\text{fwd}}w$. \\[0.2cm]
$\bullet$ Backward-forward-difference of $w$ in $y$-direction:
\begin{align} 
 \delta_y^{\text{bwd}}(\delta_y^{\text{fwd}}w) 
 \approx &  Q_y^{\text{bwd}}f_{w_y}^{
 \text{fwd}}
 \notag \\
 & -\frac{1}{h} \left[\frac{1}{h}(
  Q_{b}f_w-Q_{\mc{C}_b}f_{\psi})\right]
 \,, \label{wyycorr}
\end{align}
where  $Q_{\mc{C}_b}$ is defined in 
Eq.~(\ref{mpocb})  and $Q_{b}$ 
extracts the values of 
a function on the line $k^y = 0$,
\begin{subequations}
\begin{align}
 [Q_{b}f]_{p,q}=f_{p,q}\delta_{q,0}\,,
\end{align}
with
\begin{align}
 \lb & = 1\,, \\
 \md[k] & = \sigma_{00}\,, \quad 1\le k \le N\,, \\
 \md[k] & = \mathds{1}\,, \quad N< k\le 2N\,, \\
 \rb & = 1\,. 
\end{align}
\end{subequations}
\subsection{Laplace operator\label{laplace}}
The Laplace operator appearing in the Poisson 
equation~(\ref{poisson}) is represented 
by an MPO with bond dimension $D=6$, 
\begin{subequations}
\begin{align}
 [Q_{\Delta}f]_{p,q}=\frac{f_{p+1,q}+f_{p-1,q}
 +f_{p,q+1}+f_{p,q-1}-4 f_{p,q}}{h^2}\,,
\end{align}
with
\begin{align}
 \lb & = \left( 1, 0,0 ,1, 0,0\right)/h^2\,, \\
 \md[k] & = \left(
 \begin{array}{cccccc}
   \mathds{1} & \sigma_{01} & \sigma_{10} & 0 & 0  & 0  \\
   0 & \sigma_{10} & 0 & 0 & 0 & 0 \\
0 & 0 & \sigma_{01} &0 &0 &0 \\
0 & 0 & 0 &\mathds{1} &0 &0 \\
0 & 0 & 0 &0 &\mathds{1} &0 \\
0 & 0 & 0 &0 &0 &\mathds{1}
\end{array}
\right)\,, \quad 1\le k\le N\,, \\
 \md[k] & = \left(
 \begin{array}{cccccc}
\mathds{1} &0 &0 &0 & 0 & 0 \\
0 &\mathds{1} &0 &0 & 0 & 0 \\
0 &0 &\mathds{1} &0 & 0 & 0 \\
0 & 0  & 0 & \mathds{1} & \sigma_{01} & \sigma_{10}   \\
0 & 0 & 0 & 0 & \sigma_{10} & 0  \\
0 &0 &0 & 0 & 0 & \sigma_{01}  \\
\end{array}
\right)\,, \quad N< k \le 2N\,, \\
 \rb & = \left( -2, 1,1,-2,1,1\right)^t\,. 
\end{align}
\end{subequations}
\subsection{Central differences 
\label{mpocentral}}
Here we provide the MPO representations for the central differences in Eq.~(\ref{dxy}).

$\bullet$ Second-order accurate approximation of the first derivative in $x$-direction:
\begin{subequations}
\begin{align}
 [Q_{\partial_x}f]_{p,q} = 
\frac{1}{2h}(f_{p+1,q}-f_{p-1,q})\,, 
\end{align}
with 
\begin{align}
  \lb & = \left( 1/2, 0,0\right)/h\,, \\
 \md[k] & = \left(
 \begin{array}{ccc}
   \mathds{1} & 0 & 0 \\
   0 & \mathds{1} & 0  \\
0 & 0 & \mathds{1}  
\end{array}
\right)\,, \quad 1\le k\le N\,, \\
 \md[k] & = \left(
 \begin{array}{ccc}
   \mathds{1} & \sigma_{01} & \sigma_{10} \\
   0 & \sigma_{10} & 0 \\
0 & 0 & \sigma_{01} 
\end{array}
\right)\,, \quad N< k\le 2N\,, \\
 \rb & = \left( 0, 1,-1\right)^t\,. 
\end{align}
\end{subequations}
$\bullet$ Second-order accurate approximation of the first derivative  in $y$-direction:
\begin{subequations}
\begin{align}
 [Q_{\partial_y}f]_{p,q} = 
\frac{1}{2h}(f_{p,q+1}-f_{p,q-1})\,, 
\end{align}
with 
\begin{align}
 \lb & = \left( 1/2, 0,0 \right)/h\,, \\
 \md[k] & = \left(
 \begin{array}{ccc}
   \mathds{1} & \sigma_{01} & \sigma_{10}\\
   0 & \sigma_{10} & 0  \\
0 & 0 & \sigma_{01}  
\end{array}
\right)\,, \quad 1\le k\le N\,, \\
\md[k] & = \left(
 \begin{array}{ccc}
   \mathds{1} & 0 & 0 \\
   0 & \mathds{1} & 0  \\
0 & 0 & \mathds{1}  
\end{array}
\right)\,, \quad N< k\le 2N\,, \\
\rb & = \left( 0, 1,-1\right)^t\,. 
\end{align}
\end{subequations}
%
%
%
%
\end{document}